\documentclass[showpacs,reprint,superscriptaddress, aps, prl, floatfix]{revtex4-2}
\usepackage[utf8]{inputenc}
\usepackage{graphicx, xcolor}
\usepackage{dcolumn}
\usepackage{bm}
\usepackage{amsmath,amsthm,amssymb,bbold}
\usepackage{color}
\usepackage{verbatim}
\usepackage{physics}
\usepackage{nicefrac}
\usepackage{float}
\usepackage{natbib}
\usepackage{wasysym}
\usepackage{amsfonts}
\usepackage{booktabs}
\usepackage{times}
\usepackage{siunitx}
\usepackage[T1]{fontenc}
\usepackage{tikz}
\usepackage[export]{adjustbox}
\usepackage{tabularx}
\usepackage{multirow}

\newcommand{\Rho}{\mathrm{P}}
\newcommand\vecE{\vec{E}}
\newcommand\vecP{\vec{P}}
\newcommand\vecB{\vec{B}}
\newcommand\vecH{\vec{H}}
\newcommand\vecM{\vec{M}}
\newcommand\vece{\vec{e}}
\newcommand\vecb{\vec{b}}
\newcommand\vecA{\vec{A}}

\newcommand\rhoE{\Rho^e}
\newcommand\rhoB{\Rho^m}
\newcommand\rhoe{\rho^e}
\newcommand\rhob{\rho^m}

\newcommand{\unue}[1]{\textit{#1}---}

\definecolor{darkGreen}{RGB}{0,110,0}
\definecolor{darkBlue}{RGB}{0,0,130}
\usepackage[colorlinks,citecolor=darkGreen,linkcolor=darkBlue,urlcolor=blue,hyperindex]{hyperref}

\begin{document}
\title{Hybrid Dyons, inverted Lorentz force and magnetic Nernst effect in quantum spin ice}

\author{Chris R. Laumann}
\affiliation{Department of Physics, Boston University, Boston, Massachusetts 02215, USA}

\author{Roderich Moessner}
\affiliation{Max-Planck-Institut f\"{u}r Physik komplexer Systeme, 01187 Dresden, Germany}

\date{\today}

\begin{abstract}
Topological magnets host two sets of gauge fields: that of native Maxwell electromagnetism, thanks to the magnetic dipole moment of its constituent microscopic moments; and that of the emergent gauge theory describing the topological phase. 
Here, we show that in quantum spin ice, the emergent magnetic charges of the latter carry native electric charge of the former. 
We both provide a general symmetry-based analysis underpinning this result, and discuss a microscopic mechanism which binds a native electric charge to the emergent magnetic one. 
This has important ramifications. 
First and foremost, an applied electric field gives rise to an emergent magnetic field. 
This in turn exerts an `inverted' Lorentz force on moving emergent electric/native magnetic charges. 
This can be probed via what we term a magnetic Nernst effect: applying an electric field perpendicular to a temperature gradient yields a magnetisation perpendicular to both. 
Finally, and importantly as a further potential experimental signature, a thermal gas of emergent magnetic charges will make an activated contribution to the optical conductivity at low temperatures.
\end{abstract}

\maketitle
Spin liquids are a prominent class of model systems for the study of interacting topological phases~\cite{moessner_moore_2021}. They exhibit fractionalised excitations -- holons, spinons, monopoles -- charged under the emergent gauge field which appears in their low-energy description. Properties of these gauge fields, in particular their  signatures in experiments on candidate spin liquid materials, are a central subject of study in condensed matter and materials physics~\cite{Takagi_rev__2019,Knolle_rev_2019,udagawa2021spin}.  

In particular, how electromagnetic fields couple to emergent fractionalised degrees of freedom~\cite{Rajaraman_2001} is of both fundamental conceptual and practical importance. 
Conceptually, the question is which `quantum numbers' (or rather, charges and moments) the emergent degrees of freedom inherit from the microscopic constituents as they break apart; and practically, as the most straightforward way to couple to material noninvasively -- and hence to detect the emergent particles -- is to apply external electric and magnetic fields.  

Perhaps the simplest candidate topological quantum magnet is quantum spin ice (QSI). QSI is a term applied to a family of model systems (and materials) based on an Ising magnet on the pyrochlore lattice \cite{Anderson1956} endowed with quantum dynamics in the form of various types of spin flip terms \cite{Bramwell2001,Hermele2004,Castelnovo2012,Mcclarty2014,Rau2019}.  
\begin{figure}[tb]
    \centering
    \includegraphics[width=\columnwidth]{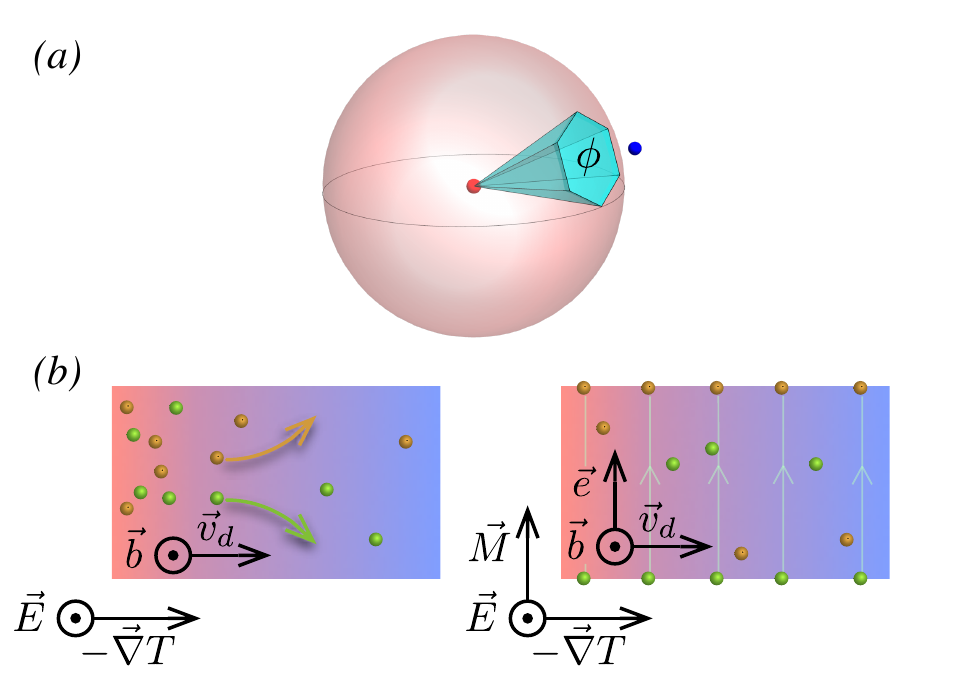}
    \caption{
    a) Emergent magnetic charge (small red sphere) emits a uniform emergent $\vec{b}$-flux. Resonance of hexagonal plaquette, Eq.~\ref{eq:ringexchdyon}, picks up a phase $\phi$, half the solid angle subtended by the plaquette, as indicated by the cone. 
    This couples linearly to the local native electric polarization, indicated by the outward displacement of an ion (blue sphere). 
    b) Left: Inverted Lorentz force--a horizontal thermal gradient $\vec{\nabla}T$ sets up a density gradient, and hence net drift current of emergent $e$-charges (green and yellow denoting opposite charges) from the hot to the cold end, where the charges get pair-created and -annihilated, respectively. 
    Switching on a native electric field, $\vec{E}$, induces a parallel emergent magnetic field, $\vec{b}$, which leads to a Lorentz force deflecting oppositely charged $e$-charges, moving with the same drift velocity $v_d$, in opposite directions, indicated by the green(yellow) arrow. 
    Right: Magnetic Nernst effect in the steady state--a surface $e$-charge distribution, which corresponds to a net magnetisation (thin arrows) $\vec{M}\propto\vec{E}\times\vec{\nabla}T$, sets up an {emergent} electric field that balances the inverted Lorentz force.
    }
    \label{fig:MagCh}
    \vspace{-\baselineskip}
\end{figure}

The gauge theory describing QSI  is an emergent form of quantum electrodynamics (eQED)~\cite{Hermele2004,MS_rvb3_2003}. It does, however, differ from the well-known native QED (nQED) believed to describe electromagnetism in matter in our universe in several crucial respects. First, its coupling strength as parametrised by its emergent fine structure constant $\alpha_{\rm e}$, is large, $\alpha_{\rm e}\agt 10 \alpha$~\cite{Pace_2021}, while its speed of emergent light, $c_e$, is much  smaller than the speed of light $c$ in vacuo~\cite{Benton2012}. 

Second, it hosts two types of emergent charges, electric and magnetic. 
In standard gauge theory language, electric fields inhabit the lattice links. Their source are charges on the lattice sites -- the emergent electric charges which arise from magnetic moment fractionalisation~\cite{Castelnovo_2008}. 
In  classical spin ice, these emergent electric charges are called magnetic monopoles as they bind an irrational~\cite{MS_irrat} native magnetic charge, $Q^{m}=2\mu/a_d$ (as well as an electric dipole moment~\cite{Khomskii_2012}) with a concomitant magnetic Coulomb interaction. 
Here, $\mu$ is the native magnetic dipole moment of the spins, while $a_d$ is a lattice constant. 

Here and in the following, we use capital letters for the native (also referred to as applied) fields ($\vec{E}, \vec{B}$) and their respective charges ($Q^e, Q^m$), with small letters denoting the corresponding emergent quantities ($\vec{e}, \vec{b}, q^e, q^m$), see Tab.~\ref{tab:fieldprops}. For the various constants, such as $\alpha$ and $c$ above, we use the subscript $_e$ to distinguish the  quantities in the emergent gauge theory. 

\begin{table}[t]
\renewcommand{\arraystretch}{1.2}
\setlength{\tabcolsep}{3pt}
    \centering
    \begin{tabular}{cc|cc|c}
        &Field & Inversion & Time-reversal &Charge (density)\\
        \cline{2-5} 
        \multirow{2}{*}{Native} & $\vec{E}$\rule{0pt}{1.2\normalbaselineskip} & $-1$ & $+1$& $Q^e \ \ (\Rho^e)$\\
        &$\vec{B}$ & $+1$ & $-1$ & $Q^m \ \  (\Rho^m)$ \\[0.5ex]
        \cline{2-5}
        \multirow{2}{*}{Emergent} & $\vec{e}$\rule{0pt}{1.2\normalbaselineskip}& $+1$ & $-1$ & $q^e \ \ (\rho^e)$\\
        &$\vec{b}$ & $-1$ & $+1$ & $q^m \ \ (\rho^m)$
    \end{tabular}
    \caption{Discrete symmetry properties of the native and emergent electromagnetic fields, and notation for their corresponding charge (densities).}
    \label{tab:fieldprops}
\end{table}

The emergent magnetic charges, $q^m$, as well as the interactions between applied electromagnetic fields and emergent electric and magnetic charges, are the focus of the present work. 
The emergent magnetic charges ($b$-charges) are not as easily visualised as their electric counterparts ($e$-charges): 
while the latter correspond to  violations of the ice rules (described below), the former manifest themselves in non-trivial phase relations in  a quantum wavefunction consisting of a superposition of many classical spin ice configurations. 

The $b$-charges therefore only emerge 
in a regime where the quantum dynamics is sufficiently coherent to allow for resonance processes between configurations involving many spins --- two spin ice configurations minimally differ by six spins arranged head-to-tail on a hexagonal loop. As illustrated in Fig.~\ref{fig:MagCh}, if the wavefunction components differing by the orientation of these six spins have a relative phase factor  
$\exp{i\phi/2}|\includegraphics[width=0.42cm,valign=c]{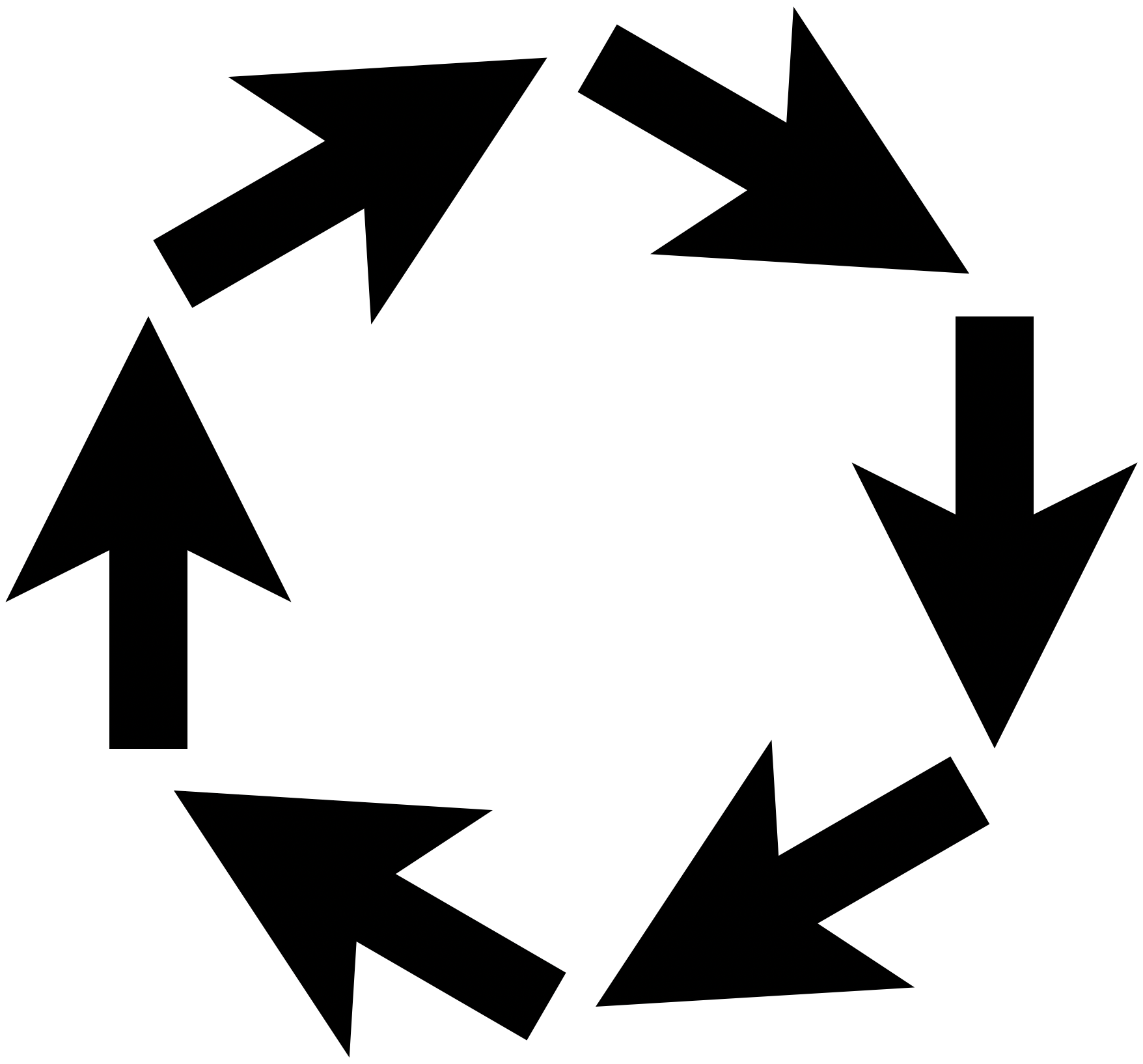}\rangle+\exp{-i\phi/2}|\includegraphics[width=0.42cm,valign=c]{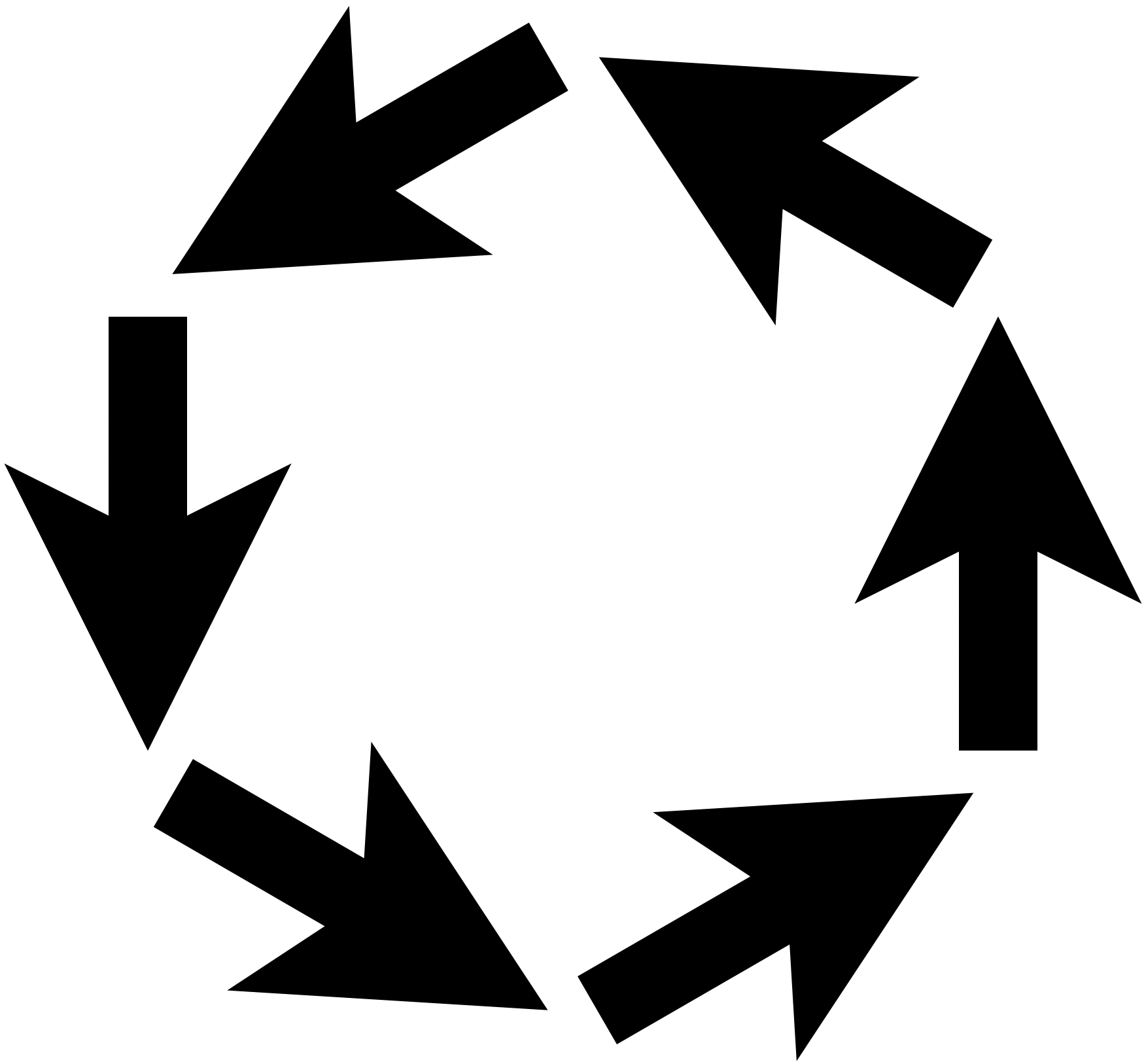}\rangle$,
this corresponds to an emergent flux $\phi$. Generally,  two configurations differing by a loop of flipped spins enter with  a relative phase proportional to the solid angle subtended by the loop with respect to the emergent magnetic charge. 
\emph{A priori,} detecting this charge requires measuring such a phase, corresponding to a high-order correlator in terms of the original spins.

Here, we provide an analysis -- long-wavelength/symmetry-based and then microscopic -- of these two sets of coupled gauge fields and  charges. 
This  allows us to address all the points raised above concerning `quantum numbers' and  coupling/detection of the emergent charges. The answers turn out to be intriguing and hold a number of surprises. 

First, an emergent magnetic $b$-charge $q^m$ also binds a native electric $E$-charge $Q^e$. 
It is in this sense that we use the word `hybrid dyon', as the combination of charges comes from marrying the two gauge theories, native and emergent. 
From this perspective, the well-known emergent electric $e$-charge is also a dyon, as it binds native magnetic charge. 
The general symmetry analysis that underlies this effect suggests that hybrid dyons are a common feature of coupled gauge theories (see e.g.~\cite{Pace2023}).

Second, we map out the response to an applied electric field, $\vec{E}$. 
This turns out to induce a uniform emergent magnetic field, $\vec{b}$, and hence exerts an `inverted' Lorentz force on the moving $e$-charge. 
As a consequence, we predict a `magnetic Nernst(-Ettinghausen) effect', where a temperature gradient orthogonal to an applied electric field goes along with a magnetisation perpendicular to both. 

Along the way, we point out that for the detection of the emergent magnetic $b$-charges,  measuring phases of an entangled wavefunction may not be necessary if one instead  directly probes the bound native  electric charge $Q^e$. 
This may for instance be done via the optical conductivity, which should contain an activated term at low temperatures, with a coefficient of the corresponding Arrhenius law given by the excitation energy of the emergent magnetic $b$-charge. 

{To link the long-wavelength considerations with   microscopic considerations, we provide a simple, detailed and transparent description of a process in a lattice model underpinning the coupling between the two gauge structures.}

\unue{Effective theory and symmetry-allowed couplings:} 
Formally, the Lagrangians  of the two gauge theories are the same ($\hbar = e = e_{e} = 1$, see Supp.~Mat. for  discussion of units):
\begin{equation}
\label{eq:twomaxwell}
    \mathcal{L} = \frac{1}{8\pi\alpha}\left(\frac{1}{c}\vecE^2 - c\vecB^2\right) + \frac{1}{8\pi\alpha_e}\left(\frac{1}{c_e}\vece^2 - c_{e}\vecb^2\right)
\end{equation}
This needs to be supplemented by the Gauss' law identifying the charge densities $\Rho,\rho$ which act as sources of the native and emergent fields, respectively. 
It is most tempting to write down the equations $\nabla\cdot\vecE=\rhoE/\epsilon_0$, $\nabla\cdot\vecB=\rhoB$, $\nabla\cdot\vece=\rhoe/\epsilon_{0e}$, $\nabla\cdot\vecb=\rhob$. 
The  formal similarity of these equations hides  important differences between native and emergent gauge theories. 
Immediately  obvious is that Maxwell's equations state that there are no sources of the magnetic field: $\rhoB\equiv 0$, 
whereas there is no such restriction on the emergent magnetic charge: $\rhob\neq0$ in general. Nonetheless, as we will expand on below, the possibility of bound charges, in the standard parlance of Maxwell electromagnetism, can at least partially plug this gap.  

The differences extend further to the symmetry properties of the fields. 
In Maxwell electromagnetism, the electric field $\vec{E}$ is a vector, while the magnetic field $\vec{B}$ is a pseudovector.
While the former is even under time reversal and odd under parity, the latter is the converse. 
By contrast, the emergent electric field $\vec{e}$ corresponds microscopically to the magnetic moment in spin ice -- and hence has the same symmetry properties as the native magnetic field, and again the converse for the emergent magnetic field. 
Linear couplings between native and emergent fields of the form $\vecb\cdot\vecB$ and $\vece\cdot\vecE$ are hence forbidden. 
Rather, allowed couplings are
\begin{equation}
\label{eq:eBbEcoupling}
    {\cal L}_{eB} = g_{eB}\, \vece\cdot\vecB\, ; \  \ {\cal L}_{bE} = g_{bE}\, \vecb \cdot \vecE\, .
\end{equation}
The first coupling is well-known.
It indicates that the emergent electric field plays the role of native magnetization in spin ice, $\vec{M} = g_{eB} \vec{e}$. 
Since a unit $e$-charge binds a magnetization monopole charge $Q^m$, as described above \cite{Castelnovo_2008}, we can fix $g_{eB} \approx \frac{2\mu}{a_d} \frac{1}{\alpha_e c_e}$. 
Likewise, an applied $\vec{B}$ field induces $\vec{e}$ which in turn yields a longitudinal force on $e$-charges. 

The second, lesser-known (though see \cite{Nakosai2019MonopoleSuppercurrent}), coupling implies that the emergent $\vec{b}$ induces an electric polarization, $\vec{P} = g_{bE} \vec{b}$. 
Furthermore, an applied electric field $\vec{E}$ induces an emergent $\vec{b}$, which imposes a Lorentz force on moving $e$-charges. 
It also implies that $b$-monopoles carry bound $E$-charge, $Q^E = 2\pi g_{bE}$.  
This underpins the striking phenoma presented in this work: the hybrid dyonic character of the emergent magnetic $b$-charges, the inverted Lorentz force, and the magnetic Nernst effect. 
We return to microscopic estimates of $g_{bE}$ below.

\unue{General considerations} One of the challenges of the present work is to make contact between different pictures that may be familiar from, e.g., undergraduate electromagnetism or graduate topological physics courses, both of which come with conceptual frameworks of their own, which are not obviously compatible. 
The key link is provided by the concepts of macroscopic bound charge and current. 
These are encoded in standard macroscopic electromagnetism in terms of the vector polarization $\vec{P}$ (not to be confused with the scalar charge densities $\Rho^{e,m}$) and magnetization $\vec{M}$ densities.

In the standard treatment, the bound charge $\Rho^e = -\nabla\cdot\vec{P}$, given by the divergence of the polarization, corresponds to an actual electric charge density.
Accordingly, in a monopole configuration, $\vec{P} = -Q^e \frac{\hat{r}}{4\pi r^2}$, the charge bound at the origin need not be quantized.
The analogous bound magnetic charge $\Rho^m = - \nabla \cdot \vec{M}$ is not typically defined in textbooks, but it is permitted so long as $\nabla\cdot\vecH=-\nabla\cdot\vecM$ \cite{MS_irrat}.
As far as the response of the medium goes, whether or not either type of charge is free or bound need not play a large role, but it is not immaterial either. 
For instance,  DC currents of bound charges do not exist, as these would imply an unbounded build-up of polarisation or magnetization.

\unue{Hybrid dyons and their currents} 
As noted above, emergent $b$-charges bind irrational native $E$-charge, just as emergent $e$-charges bind native magnetization $B$-charge. 
Unlike `proper' dyons, there is no Dirac-Zwanziger \cite{Zwanziger1968} quantization condition governing the amount of bound native charge on an emergent excitation, nor do P and T-breaking perturbations lead to a Witten effect \cite{Witten1979Dyon}, as they do for the allowed emergent $e$- and $b$-charges \cite{Pace2021b}.

An applied electric $\vec{E}$ or magnetic $\vec{B}$ field can thus induce emergent currents. In particular, Eq.~\ref{eq:eBbEcoupling} implies that $\vec{E}$ will drive a current of emergent magnetic $b$-charges. The complement, $\vec{B}$ inducing a current of emergent electric $e$-charges, has been studied under the heading of magnetricity~\cite{magnetricity,runaway}. 

\unue{Activated optical conductivity}
While an emergent electric current induces a change in the native magnetisation of the system, the emergent magnetic current changes the native electric polarisation. An applied electric field inducing a polarisation may not be particularly surprising, just as magnetricity describes  an applied magnetic field magnetising a sample. The challenge is how to separate out the contribution of the topological magnetism to the total induced polarisation. 

Most simply, the density of emergent magnetic $b$-charge carriers is activated with an Arrhenius-type activation gap set by the (effective) ring-exchange energy scale $g_r$ in Eq.~\ref{eq:ringexchdyon}. 
As these carriers carry native $E$-charge, they tend to screen applied $E$-fields much like a compensated semiconductor.

To detect these charges, one  therefore needs to separate out the relevant activated low-temperature contribution from a measurement of the dielectric constant of the material as a function of temperature. In practise, a number of conditions have to be met. 
First, the measurement has to be conducted at a  frequency, $\omega$,  high enough for there to  be no significant accumulation of bound charge on the sample surface. This amounts to $\rhob v_d a_d^2/\omega\ll 1$, where $v_d$ is the drift velocity of the charges.  
At the same time, the frequency needs to be low enough for the $b$-charge carriers to respond as well-defined carriers -- a scale set by ring exchange $g_r$. 

Given this ring-exchange scale $g_r$ is typically believed to be a rather small, in the sub-Kelvin regime, most phononic modes will be well frozen out, but distinguishing  `non-topological' background, e.g.\ due to possible impurities, from the signal will presumably nonetheless be a challenge. 

\unue{Inverted Lorentz force}  The  emergent fields generated through an applied electro-magnetic field via the linear couplings in Eq.~\ref{eq:eBbEcoupling} can have further observable effects. The possibility of coupling an applied electric field to an emergent electric $e$-charge via the latter's native electric dipole moment has been noted before~\cite{Khomskii_2012}, as has the coupling of an applied magnetic field to an emergent magnetic $b$-charge \cite{Gang_Hall}.

Here, we introduce a new, `inverted', Hall response:  applying uniform $\vecE$  induces a uniform $\vecb$, yielding a  Lorentz force on a moving emergent electric charge, Fig.~\ref{fig:MagCh}(b).  
On account of its hybrid dyonic character, a (Hall) current of $\rhoe$  thus goes along with one of $\rhoB$. The latter simply amounts to an accumulation of (bound) charge on a sample surface, and hence a net {\it magnetisation} perpendicular to the applied electric field. We call this an inverted Hall effect, as an applied {\it electric} field yields a (native) {\it magnetic} response. 

Now, there is an obstacle to the observation of Hall currents of this type -- in spin ice, positive and negative $e$-charges are  perfectly compensated (to borrow the term from semiconductor physics). Their respective Hall currents therefore cancel.

This cancellation is not the end of story, however. In the context of electronic physics, the study of various longitudinal and transverse transport coefficients has a long and distinguished history, a prominent subject of which has been the (longitudinal) Seebeck and (transverse) Nernst(-Ettinghausen) effects. It is the latter -- observable also in compensated situations -- that we therefore turn  our attention to.

\unue{Magnetic Nernst effect} 
The missing ingredient is a thermal gradient $\vec\nabla T$. 
This generates a density gradient of $e$-charges, as there is a higher density of activated charges on the hotter side of the sample. This leads to a net particle (but zero net charge) current from the hot to the cold region, with the local equilibrium densities being established by pair creation (annihilation) in the hotter (colder) regions, see Fig.~\ref{fig:MagCh}b). 

When an electric field $\vec{E}$ is applied perpendicularly to the temperature gradient, the emergent electric $e$-charges  experience the inverted Lorentz force. 
Crucially, this points in opposite directions for the opposite emergent electric charges, as their drift velocity along the temperature gradient points in the same direction. 
The  results is a net emergent electric charge current, perpendicular to both the temperature gradient and the applied electric field. This current results in the build-up of a magnetisation, for more details see Fig.~\ref{fig:MagCh}. 

This combination  of perpendicular temperature gradient, applied electric field, and induced magnetisation, is what we call the magnetic Nernst effect. 
This is in contrast to the conventional Nernst effect where an electric field arises in the presence of an applied magnetic field and thermal gradient.

\unue{Microscopic picture}
The reader may  worry that having an effective theory is not quite sufficient as a basis for the far-reaching results we have presented. It is at any rate clearly desirable to analyse a microscopic model exhibiting these effects as a point of principle. 
This can also yield  an idea on which (physically measurable) quantities the coupling $g_{bE}$ depends. 

To provide this, we proceed in two steps. We first identify a microscopic operator with the required symmetry properties, and then present a toy model to make transparent a mechanism by which this term arises. 

Candidate QSI materials are magnetic insulators in the pyrochlore family with generic chemical formula $A_2 B_2$O$_7$. Here $A$ and $B$ are usually rare-earth and transition metal ions, respectively \cite{Gardnner2010,Ross2011,YbTO_guidi,CZO_mc,CZO_gao,PHO_lake,PHO_psi,PZO_spinorbital,ice_ultra}, residing on interpenetrating pyrochlore lattices of corner-sharing tetrahedra. 
The local [111] axis, a three-fold rotational symmetry axis, is a natural quantisation direction. The local ground state doublet is then parametrised by the direction of the (pseudo-)spin of the $A$-ion along this axis, represented with the Pauli operators $\vec{S}_i$, with the convention that $S^z_i = \pm 1$ if the magnetic moment at atom $i$ points from an up-pointing tetrahedron to a down-pointing one. 

When the interaction Hamiltonian is dominated by an effectively  ferromagnetic nearest-neighbor Ising coupling, $H = J_{zz} \sum_{\langle i j\rangle} S^z_i S^z_j$, the many-ion ground states satisfy the ice rule: 
at any tetrahedron, two moments point in and two point out. 
In more formal terms, this is an emergent lattice Gauss' law, $\mathrm{div}_t S^z = 0$ where $t$ runs over tetrahedra.
If  off-diagonal contributions to $H$ are so weak that they only violate the ice rule virtually, one may project the residual couplings into the ice manifold to obtain an effective ring exchange Hamiltonian
\begin{align}
    H_{ice} = - g_{r} \sum_p S^+_{p_1} S^-_{p_2} S^+_{p_3} S^-_{p_4} S^+_{p_5} S^-_{p_6} + \emph{h.c.} \ .
    \label{eq:ringexchdyon}
\end{align}
Here, $p$ runs over all of the hexagonal plaquettes in the pyrochlore lattice and $g_r$ is the ring exchange coupling. 
$H_{ice}$ is the leading symmetry allowed term which does not violate the ice rule, as it only acts to reverse cycles of moments pointed head-to-tail; pictorially, it may be represented $|\includegraphics[width=0.42cm,valign=c]{ccw.png} \rangle\langle \includegraphics[width=0.42cm,valign=c]{cw.png}|$. 
Furthermore, $H_{ice}$ realizes a deconfined Coulomb phase, in which the emergent electric field $e_i$ is identified with the local $S^z_i$ moment and the conjugate vector potential $a_i$ to the off-diagonal raising and lower operators $S^\pm = e^{\pm i a_i}$ (see \cite{Hermele2004}).
In this language, $H_{ice} = - 2 g_r \sum_p \cos (\mathrm{curl}_p a)$. 

Any microscopic operator which transforms like a polar, T-even vector under the point group of the pyrochlore crystal is allowed to contribute to the (native) electric dipole moment $\vec{d}$ and couple to $\vec{E}$ linearly. 
From the long-wavelength considerations, we expect the microscopic counterpart to the magnetic flux $\vec{b} \approx \vec{\nabla} \times \vec{a}$ to behave appropriately. 
In the lattice gauge theory rewriting, the emergent magnetic flux through a hexagonal plaquette is $b_p = \mathrm{curl}_p a \approx \sin(\mathrm{curl}_p a)$.
This leads us to consider the dimensionless microscopic spin operator
\begin{align}
\label{eq:micro_b_polarization}
	\vec{O} = i \sum_{p} \hat{t}_p \, S^+_{p_1} S^-_{p_2} S^+_{p_3} S^-_{p_4} S^+_{p_5} S^-_{p_6}
\end{align}
where $p$ runs over all oriented hexagonal plaquettes in the pyrochlore lattice and $\hat{t}_p$ is the unit vector `normal' to the plaquette with sign given by the right-hand rule. 
It is now an algebraic exercise to verify that $\vec{O}$ indeed transforms as a T-even vector under the point group of the pyrochlore for all of the symmetry classes of quantum spin ice (see Supp.~Mat.\ for details).  
We expect that the 6-body operator $\vec{O}$ is the most local vector operator which survives the ice projection--at any rate, it has been shown that none are available for bilinear nearest-neighbour coupling~\cite{Lantagne-Hurtubise2017}.

Finally, we turn to an actual physical mechanism which allows estimating the coupling $\vec{P} \sim g_{bE} \vec{O}$. 
At first glance, in an insulator, it is unclear how any such coupling between magnetic moments and native charge polarization can arise. The Suppl.\ Mat.\ includes a fully-worked toy model on a cubic version of the pyrochlore lattice, the geometry of which hugely simplifies algebra and notation. The upshot is the following. 

The atomic dipole moment operators 
\begin{align}
    \vec{d}_p &= \sum_{i \in p} \vec{d}_i
\end{align}
for atoms $i$ on the edges of a plaquette $p$  can be formally projected into the ice manifold to obtain an effective plaquette dipole moment operator
\begin{align}
    \vec{d}_{eff} &= P_{ice} e^{\hat{S}_{SW}}(\vec{d}_p) P_{ice} \ .
\end{align}
Here, $\vec{d}_i$ is the dipole moment of the electron cloud relative to the center of its atom $i$. 
Parity selection rules require $\vec{d}_i$ to vanish within the crystal field manifold of the low lying $J$-multiplet of the $A$ ion. Accordingly, the Schrieffer-Wolff projection leads to an estimate
\begin{align}
    \vec{d}_{eff} &\sim d_0 \frac{J_{NN}^3}{\Delta_{L} \Delta_{ice}^2} \hat{O}
\end{align}
where $d_0$ is the characteristic scale of the dipole moment of the $A$ ion, $J_{NN}$ are nearest neighbor magnetic coupling scales, $\Delta_L$ is the energy of an electronic configurational excitation in the $A$ ion and $\Delta_{ice}$ is the gap to breaking the ice rules. 
We note that this estimate provides a minimal contribution to the polarization $\vec{P}$ due to $\vec{b}$--the true polarization may be enhanced due to additional physical processes.

\unue{Outlook} 
The prediction of the emergent $b$-charge as a stable quasiparticle in  compact QED is one of the central qualitative features of the QSI phase.
Adducing experimental evidence of these has turned out to be a formidable challenge.
Our work provides a route to experimental identification of these particles and related phenonemona associated with the 'coherence' of the quantum spin ice state.
Indeed, the role of coherence and determination of the stability of the quasiparticles, their drift velocities and the way in which the bound charges described here 'come along for the ride' are all  interesting topics in their own right. 
Clearly, the frequency and temperature scales required to observe the effects described here depend on these quantities, whose detailed determination is not a simple theoretical exercise. 
We hope that near-term experiments will shed light on these fascinating issues. Beyond magnetic materials, one may also dream of artifical implementations of spin ice physics~\cite{Colloidal_ice,ASI_review_nisoli,ASI_review_stamps,QubitSI,Shah2023QSIRydberg}, where the role of the native fields may differ qualitatively on account of the different nature of the underlying microscopic degrees of freedom.

\begin{acknowledgments}
\textbf{Acknowledgements:}
The authors are grateful to Subhro Bhattacharjee, Claudio Castlenovo, Anushya Chandran, and Salvatore Pace for discussions. 
This work was in part supported by the Deutsche Forschungsgemeinschaft  under grants SFB 1143 (project-id 247310070) and the cluster of excellence ct.qmat (EXC 2147, project-id 390858490).  
C.R.L. acknowledges support from the NSF through grant PHY-1752727. 
\end{acknowledgments}
\bibliographystyle{apsrev4-1}
\bibliography{references}

\pagebreak
\onecolumngrid
\pagebreak

\section*{Supplemental Material}

\section{Microscopic Polarization} 
\label{sec:microscopic_polarization}

Any operator which transforms like a polar, T-even vector under the point group of the pyrochlore crystal should, on symmetry grounds, contribute to the polarization $\vec{P}$ (ie. couple to $\vec{E}$ linearly). 
Ref.~\cite{Lantagne-Hurtubise2017} found all vector-like irreps of the diamond lattice point-group up to bilinear in nearest-neighbor spin operators; projecting these down into the spin ice manifold leads to quadratic coupling to $\vec{E}$. From the long-wavelength considerations, we expect a linear coupling to exist microscopically, but it will involve more than nearest neighbor spin operators and an exhaustive analysis would be exhausting. Rather, here we simply guess: the $\vec{b}$ corresponds to the flux of the lattice $\vec{a}$, so it should have something to do with oriented ring exchange coupled to a vector given by the right hand rule.

\begin{widetext}
We define the ring raising operator
	\begin{align}
	\Lambda_{ik}^{j, s} = 
	S^+_{s + (-1)^s (\vec{t}_i/2)} 
	S^-_{s + (-1)^s (\vec{t}_i - \vec{t}_j/2)} 
	S^+_{s + (-1)^s (\vec{t}_i - \vec{t}_j+\vec{t}_k/2)} 
	S^-_{s + (-1)^s (\vec{t}_i/2 - \vec{t}_j+\vec{t}_k) } 
	S^+_{s + (-1)^s (- \vec{t}_j/2 + \vec{t}_k)}
	S^-_{s + (-1)^s (\vec{t}_k/2)}
\end{align}
\end{widetext}
which raises the circulation of the spins around the elementary cycle specified by  $(s,i\bar{j}k\bar{i}j\bar{k})$ for site $s\in A$ or $(s, \bar{i} j \bar{k} i \bar{j} k)$ for $s \in B$ with distinct $i,j,k \in \{1\cdots 4\}$ indexing the 4 $[111]$ directions in the lattice. 
The hermitian conjugate of $\Lambda$ lowers the circulation, or equivalently, raises the circulation of the oppositely oriented cycle:
\begin{align}
	\left(\Lambda_{ik}^{j,s}\right)^\dagger = \Lambda_{ki}^{j,s}
\end{align}
Also, the following bookkeeping identities hold,
\begin{align}
	&\underbrace{\Lambda_{ik}^{j,s} = \Lambda_{kj}^{i,s+i-j} = \Lambda_{ji}^{k,s-j+k}}_{A\textrm{ site reference}}
	\nonumber \\ &= \underbrace{(\Lambda_{ji}^{k,s+i})^\dagger = (\Lambda_{ik}^{j,s+i-j+k})^\dagger   = (\Lambda_{kj}^{i,s+k})^\dagger}_{B\textrm{ site reference}}
\end{align}

Define the average lattice flux operator
\begin{align}
\label{eq:micro_b_polarization_SI}
	\vec{O} = i \sum_{s\in A} \sum_{\sigma \in S_4} \Lambda_{\sigma_1 \sigma_3}^{\sigma_2, s} \hat{t}_{\sigma_4} \, \mathrm{sign} (\vec{t}_{\sigma_4} \cdot (\vec{t}_{\sigma_1} \times \vec{t}_{\sigma_3})) 
\end{align}
The sum on A sites automatically makes this operator invariant under lattice translations. 
The sum on permutations $\sigma$ runs over all 24 of the oriented cycles running through each A site. 
As every distinct cycle runs through 3 A sites, $\vec{O}$ triple counts the cycles, but this simply contributes a factor of 3 compared to the less formal definition in the main text.
The unit vectors $\hat{t}_i$ correspond to the 4 $[111]$ directions.

\begin{table}[t]
	\centering

	\begin{tabular}{l|cc}
	\hline

	\hline
	\textbf{Doublet} & \textbf{State} & \textbf{Number of Spins-1/2} \\
	\hline
	\textrm{Effective spin-1/2} & $\ket{\pm 1/2}$ & $n = 1$\\
	\textrm{Dipolar-octupolar}  & $\ket{\pm 3/2}$ & $n = 3$\\
	\textrm{Non-Kramers}	    & $\ket{\pm 1}$  &  $n = 2$\\
	\hline

	\hline
	\end{tabular}
	\caption{Types of crystal field doublets in pyrochlore spin ice.}
	\label{tab:spin_symmetry}
\end{table}

We next specify the behavior of the microscopic spin operators $\vec{S}_r$ under the point group and time reversal transformations. 
Quantum spin ices are grouped according to the representation of the fundamental doublet `spin' \cite{Rau2019}, see Table~\ref{tab:spin_symmetry}.
We get a nice compact representation for each of the classes by simply putting $n = 2s$ `true' spin 1/2's per site with the following definitions:
\begin{align}
	S^z &= \frac{1}{n}\sum_{j=1}^n \sigma^z_j &
	S^{\pm} &= \prod_{j=1}^n \sigma^{\pm}_j
\end{align}
From these formulae, the transformation rules for the effective spin operators follow immediately from those of a fundamental spin $\vec{\sigma}$, which is a T-odd axial vector.
These choices normalize $S^z, S^\pm$ as Pauli matrices.

\paragraph{Spin-1/2---} If the spin operators behave like fundamental spin 1/2 objects, then $\vec{S}$ is a T-odd axial vector. That is:
\begin{align}
	T: \vec{S}_r &\to -\vec{S}_r \\
	R: \vec{S}_r &\to |R|R \vec{S}_{R^{-1}r}
\end{align}

Breaking this into local components, we find time reversal is straightforward
\begin{align}
	T: S^z_r &\to - S^z_r \\
		S^\pm_r&\to - S^{\mp}_r \\
		\Lambda_{ik}^{j,s} &\to \Lambda_{ki}^{j,s} = \left(\Lambda_{ik}^{j,s}\right)^\dagger
\end{align}

\begin{widetext}
Spatial operations require the spin fields to be re-expanded in the local basis at the new bond:
\begin{align}
	\vec{S}_r = S^z_r \hat{t}_r + S^+_r \vec{e}^-_r + S^-_r \vec{e}^+_r
\end{align}
Under point group transformations $R$ which preserve $s$ sites, the local $\hat{t}_r$ vector is preserved by $R$, even if it is improper:
\begin{align}
	\hat{t}_r = R \hat{t}_{R^{-1}r}\ .
\end{align}
The transverse planes are mapped into one another by $R$ as well, but the local choice of $\hat{x}$ and $\hat{y}$ may require an extra $O(2)$ rotation to be brought into alignment:
\begin{align}
	e^{\pm i \theta(R,r)} \vec{e}^{|R|\pm}_r =  R \vec{e}^{\pm}_{R^{-1}r}
\end{align}
Note that the O(2) rotation is improper if and only if $R$ is. The angle $\theta(R,r)$ only depends on the type $i$ of bond center $r$ and the point group transformation $R$. 

Putting this together, we find that
\begin{align}
	R : \vec{S}_r &\to |R| R \vec{S}_{R^{-1}r} \\
	&= |R| R \left(S^z_{R^{-1}r}\hat{t}_{R^{-1}r} + S^+_{R^{-1}r} \vec{e}^-_{R^{-1}r} + S^-_{R^{-1}r} \vec{e}^+_{R^{-1}r} \right) \\
	&= |R| \left(S^z_{R^{-1}r} R \hat{t}_{R^{-1}r} + S^+_{R^{-1}r} R \vec{e}^-_{R^{-1}r} + S^-_{R^{-1}r} R \vec{e}^+_{R^{-1}r} \right) \\
	&= |R| \left(S^z_{R^{-1}r} \hat{t}_{r} + S^+_{R^{-1}r} e^{-i \theta(R,r)} \vec{e}^{|R|-}_{r} + S^-_{R^{-1}r} e^{+i \theta(R,r)}\vec{e}^{|R|+}_{r} \right) 
\end{align}
which allows us to extract the transformation law for the component fields:
\begin{align}
	R : S^z_{r} &\to |R| S^z_{R^{-1}r} \\
      S^{\pm}_r &\to |R| e^{\mp i \theta(R,r)}S^{|R|\pm}_{R^{-1}r}
\end{align}
and for the ring exchange operators:
\begin{align}
	\Lambda_{ik}^{j,s} &\to |R|^6 e^{i \sum \theta}
	S^{|R|+}_{R^{-1}(s + \vec{t}_i/2)}
	S^{|R|-}_{R^{-1}(s + \vec{t}_i - \vec{t}_j/2)}
	S^{|R|+}_{R^{-1}(s + \vec{t}_i - \vec{t}_j+\vec{t}_k/2)}
	S^{|R|-}_{R^{-1}(s + \vec{t}_i/2 - \vec{t}_j+\vec{t}_k)}
	S^{|R|+}_{R^{-1}(s - \vec{t}_j/2 + \vec{t}_k)}
	S^{|R|-}_{R^{-1}(s + \vec{t}_k/2)} \nonumber\\
	&= \left\{ \begin{array}{ll}
	\Lambda_{R^{-1}i R^{-1}k}^{R^{-1}j, R^{-1}s} & |R| = +1 \\
	\Lambda_{R^{-1}k R^{-1}i}^{R^{-1}j, R^{-1}s} & |R| = -1 
	\end{array} \right.
\end{align}
Here, $R^{-1}$ acts as permutation on the bond types $i$ because $R$ which preserves $s$ simply permutes the $\hat{t}_i$. More specifically, we define the permutation action of $R$ on the bond types by $R \vec{t}_i = \vec{t}_{R i}$. 

Finally, consider inversion through a bond center $I$. This maps all $i$-type bonds to $i$-type bonds. Thus:
\begin{align}
	\hat{t}_r &= \hat{t}_{I^{-1} r} &
	\hat{e}^{\pm}_r &= \hat{e}^{\pm}_{I^{-1}r}
\end{align}
So we find the trivial action on the spin components
\begin{align}
	I : \vec{S}_r &\to |I| I \vec{S}_{I^{-1} r} = \vec{S}_{I^{-1} r} \\
		S^z_r &\to S^z_{I^{-1} r} \\
		S^\pm_r & \to S^{\pm}_{I^{-1} r} 
\end{align}
On the other hand, the ring exchange orientation is flipped:
\begin{align}
	I: \Lambda_{ik}^{j,s} \to \Lambda_{ik}^{j, (I^{-1}s)} = \Lambda_{ki}^{j,I^{-1}s +(-1)^s(- i + j - k)}
\end{align}
That is, if $I$ is plaquette $p = (s, ijk)$ centered, then $\Lambda_p \to \Lambda_p^\dagger$.

Finally, we turn to the action of the symmetries on  $\vec{O}$. For proper $R$ preserving s:
\begin{align}
	R : \vec{O} &= i \sum_s \sum_\sigma \Lambda_{\sigma_1 \sigma_3}^{\sigma_2, s} \hat{t}_{\sigma_4} \, \mathrm{sign} (\vec{t}_{\sigma_4} \cdot (\vec{t}_{\sigma_1} \times \vec{t}_{\sigma_3})) \\
	&\to i \sum_s \sum_\sigma \left(\Lambda_{R^{-1} \sigma_1 R^{-1}\sigma_3}^{R^{-1}\sigma_2, R^{-1}s} \right) \hat{t}_{\sigma_4} \, \mathrm{sign} (\vec{t}_{\sigma_4} \cdot (\vec{t}_{\sigma_1} \times \vec{t}_{\sigma_3})) \\
	&= i \sum_s \sum_\sigma \Lambda_{\sigma_1 \sigma_3}^{\sigma_2, s} \hat{t}_{R \sigma_4} \, \mathrm{sign} (\vec{t}_{R \sigma_4} \cdot (\vec{t}_{R \sigma_1} \times \vec{t}_{R \sigma_3})) \\
	&= i \sum_s \sum_\sigma \Lambda_{\sigma_1 \sigma_3}^{\sigma_2, s} R \hat{t}_{\sigma_4} \, \mathrm{sign} (R \vec{t}_{\sigma_4} \cdot (R \vec{t}_{\sigma_1} \times R \vec{t}_{\sigma_3})) \\
	&= i \sum_s \sum_\sigma \Lambda_{\sigma_1 \sigma_3}^{\sigma_2, s} R \hat{t}_{\sigma_4} \, \mathrm{sign} (\vec{t}_{\sigma_4} \cdot (\vec{t}_{\sigma_1} \times \vec{t}_{\sigma_3})) = R \vec{O}
\end{align} 
For improper $R$ preserving s, 
\begin{align}
	R : \vec{O} &\to i \sum_s \sum_\sigma \left(\Lambda_{R^{-1} \sigma_3 R^{-1}\sigma_1}^{R^{-1}\sigma_2, R^{-1}s} \right) \hat{t}_{\sigma_4} \, \mathrm{sign} (\vec{t}_{\sigma_4} \cdot (\vec{t}_{\sigma_1} \times \vec{t}_{\sigma_3})) \\
	&= i \sum_s \sum_\sigma \Lambda_{\sigma_3 \sigma_1}^{\sigma_2, s} R \hat{t}_{\sigma_4} \, \mathrm{sign} (R \vec{t}_{\sigma_4} \cdot (R \vec{t}_{\sigma_1} \times R \vec{t}_{\sigma_3})) \\
	&= i \sum_s \sum_\sigma \Lambda_{\sigma_3 \sigma_1}^{\sigma_2, s} R \hat{t}_{\sigma_4} \, |R| \mathrm{sign} (\vec{t}_{\sigma_4} \cdot (\vec{t}_{\sigma_1} \times \vec{t}_{\sigma_3})) \\
	&= -i \sum_s \sum_\sigma \Lambda_{\sigma_3 \sigma_1}^{\sigma_2, s} R \hat{t}_{\sigma_4} \, |R| \mathrm{sign} (\vec{t}_{\sigma_4} \cdot (\vec{t}_{\sigma_3} \times \vec{t}_{\sigma_1})) 
	= R \vec{O}
\end{align} 
and for bond centered inversion $I$:
\begin{align}
	I : \vec{O} &\to i \sum_s \sum_\sigma \Lambda_{\sigma_3 \sigma_1}^{\sigma_2, (Is - \vec{t}_{\sigma_1} + \vec{t}_{\sigma_2} - \vec{t}_{\sigma_3})} \hat{t}_{\sigma_4} \, \mathrm{sign} (\vec{t}_{\sigma_4} \cdot (\vec{t}_{\sigma_1} \times \vec{t}_{\sigma_3})) \\
	&= i \sum_s \sum_\sigma \Lambda_{\sigma_3 \sigma_1}^{\sigma_2, s} \hat{t}_{\sigma_4} \, \mathrm{sign} (\vec{t}_{\sigma_4} \cdot (\vec{t}_{\sigma_1} \times \vec{t}_{\sigma_3})) \\
	&= - i \sum_s \sum_\sigma \Lambda_{\sigma_3 \sigma_1}^{\sigma_2, s} \hat{t}_{\sigma_4} \, \mathrm{sign} (\vec{t}_{\sigma_4} \cdot (\vec{t}_{\sigma_3} \times \vec{t}_{\sigma_1})) = I \vec{O}
\end{align}
and for time reversal $T$:
\begin{align}
	T : \vec{O} &\to -i \sum_s \sum_\sigma \Lambda_{\sigma_3 \sigma_1}^{\sigma_2, s} \hat{t}_{\sigma_4} \, \mathrm{sign} (\vec{t}_{\sigma_4} \cdot (\vec{t}_{\sigma_1} \times \vec{t}_{\sigma_3})) = \vec{O}
\end{align}
So, indeed, the $\vec{O}$ operator transforms like a T-even polar vector.

\end{widetext}

\paragraph{Other classes---} 
For $n > 1$, the spin operators transform as:
\begin{align}
	T : S^z_r &\to - S^z_r \\
		S^\pm_r &\to (-1)^n S^{\mp}_r \\
	R : S^z_r &\to |R| S^z_{R^{-1}r} \\
		S^\pm_r &\to |R|^n e^{\mp i n \theta(R,r)} S^{|R|\pm}_{R^{-1}r} \\
	I : S^z_r &\to S^z_{I^{-1}r} \\
		S^\pm_r & \to S^\pm_{I^{-1}r}
\end{align}
It is straightforward to check that the additional $n$-dependent factors cancel out of the transformation rules for $\Lambda$ and, accordingly, $\vec{O}$.

\section{Units}

In SI units, the Lagrangian for the (native) electromagnetic field in matter can be written generally
\begin{align}
    \mathcal{L} = \frac{\epsilon_0}{2}(\vecE^2 - c^2 \vecB^2) - \phi \rho_f + \vecA\cdot \vec{J}_f + \vecP \cdot \vecE + \vecM \cdot \vecB
\end{align}
where $\vecE = - \vec{\nabla} \phi - \dot{\vecA}$ and $\vecB = \vec{\nabla}\times \vecA$. 
This allows us to read off the magnetization $\vecM$ and polarization $\vecP$ by comparison with the Lagrangian of the coupled system in the main text.

The fine structure constant $\alpha = \frac{1}{4\pi \epsilon_0} \frac{e^2}{\hbar c}$ and $c = \frac{1}{\sqrt{\mu_0 \epsilon_0}}$; thus, both $\epsilon_0$  and $\mu_0$ can be expressed in terms of fundamental constants $e, \hbar, c$ and $\alpha$.
Since $c$ and $\alpha$ are different in the native and emergent theories, we find it convenient to set $\hbar = e = e_e = 1$ and parametrize the theory by $c, c_e$ and $\alpha, \alpha_e$ explicitly as in Eq.~\eqref{eq:twomaxwell}.
The fields due to elementary emergent charges are then
\begin{align}
\label{eq:elementaryfields}
    \vec{e} &= \alpha_e c_e \frac{\hat{r}}{r^2} \\
    \vec{b} &= \frac{1}{2} \frac{\hat{r}}{r^2}
\end{align}
where we have taken the elementary $b$-monopole charge $q^m=2\pi$ due to Dirac quantization,
\begin{align}
    q^e q^m & = 2\pi \hbar n & n \in \mathbb{Z}
\end{align}

Noting that $\vec{P} = g_{bE} \vec{b}$ and that the bound native charge $\rho_b = - \vec{\nabla}\cdot\vecP$, we obtain the relationship $Q^e = - 2\pi g_{bE}$ for the charge bound on an elementary $b$-monopole using Eq.~\eqref{eq:elementaryfields}.

\section{Toy Model in Octahedral Spin Ice}

The crystal field environment of a hexagonal plaquette in pyrochlore spin ice is geometrically quite complicated. 
Here, we provide a toy model in octahedral spin ice with cubic symmetry. 
This permits a simple illustration of how the atomic dipole operator on a square plaquette rotates into the effective ring exchange operator $\vec{O}$ (see Eq.~\eqref{eq:micro_b_polarization}) by Schrieffer-Wolf rotation into the low energy manifold. 

\begin{figure}[tb]
    \centering
    \includegraphics[width=0.8\columnwidth]{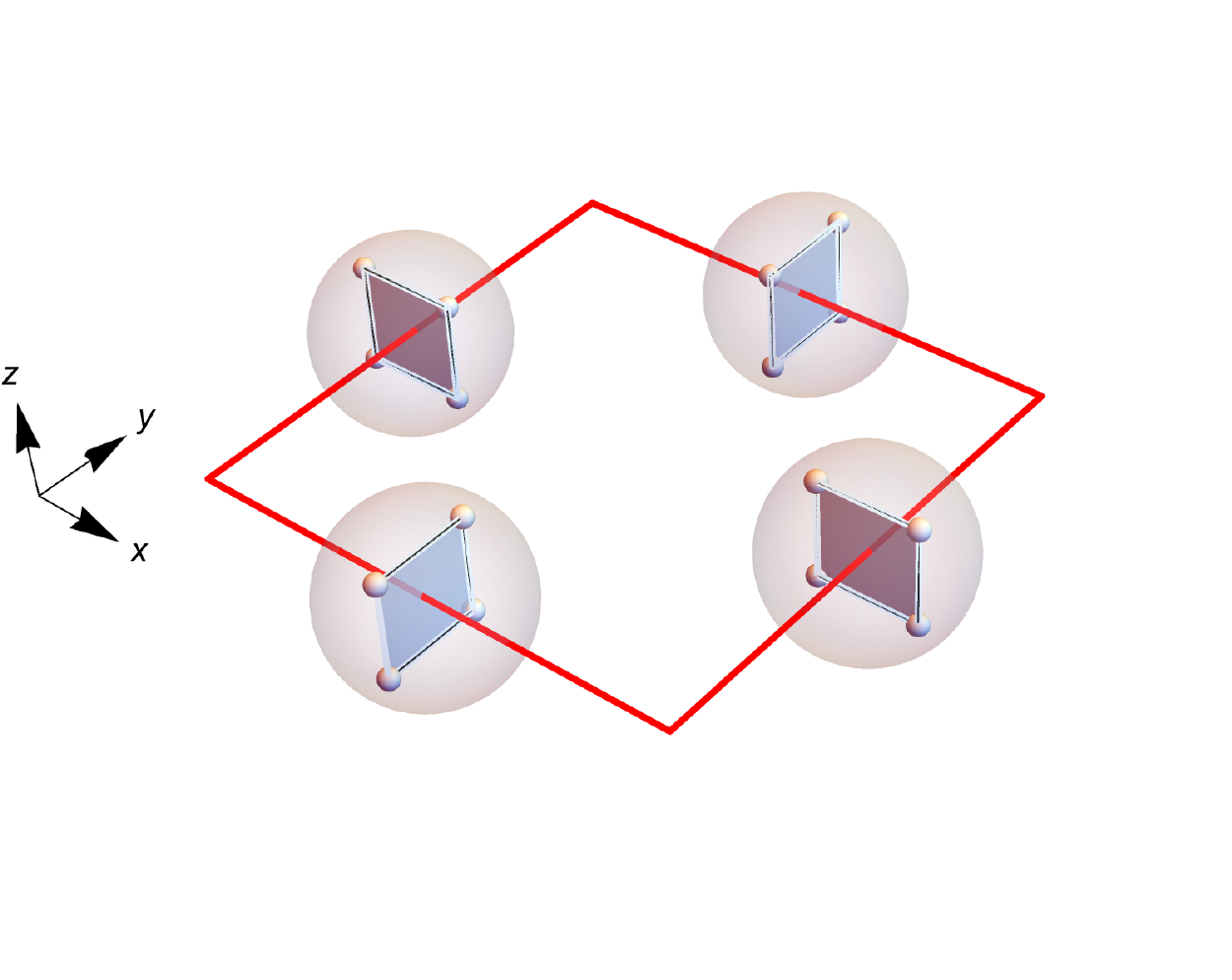}
    \caption{Representative square plaquette in toy model for octahedral spin ice. Each of the 4 atoms (transparent spheres) is located at the midpoint of an edge (red) of the plaquette. Each atom contains 4 orbitals arranged in a ring on the plane transverse to its edge, between which two spinless electrons hop. This leads to a ground state orbital doublet whose states correspond to orbital angular momentum aligned along or against the local axis.}
    \label{fig:fig_toy_square_plaq}
\end{figure}

In octahedral spin ice, the magnetic atoms sit on the midpoints of the bonds of a cubic lattice, see Fig.~\ref{fig:fig_toy_square_plaq}, and the ice rules take the form of 'three-in--three-out' configurations on each vertex of the cubic lattice. 
In the extreme crystal field dominated limit, we model each `atom' by a hopping model on a set of four `atomic' orbitals arranged in a square ring transverse to the bond. 
For example, the orbitals for an atom on an $\hat{x}$ bond are displaced from the midpoint of the bond by $l (\pm \hat{z} \pm \hat{y})/\sqrt{2}$ with some distance $l$ much smaller than the lattice length $1$. 
The intra-atom crystal field Hamiltonian on atom $i$ is simply 
\begin{align}
    H_{XF,i} = -t \sum_a c^{\dagger}_{i,a+1} c_{i,a} + \textrm{h.c.}
\end{align}
where $a$ runs over the four orbitals. 
The single particle spectrum of $H_{XF}$ is $-t, 0, 0, t$, with the central doublet carrying angular momentum $\pm \pi/2$ (clockwise/counter-clockwise) around the ring.
By placing two spinless, non-interacting electrons on each atom, we obtain an easy-axis non-Kramers ground state doublet corresponding to whether the second electron is in the CW or CCW state -- that is, whether the current is clockwise or counter-clockwise. 
The electrons cannot hop between atoms and they do not interact within the atom. 
 
The ice rules emerge from magnetic (current-current) interactions between neighboring atoms
\begin{align}
    H_{ice} &= J \sum_{\langle i,j \rangle} \sigma^z_i \sigma^z_j
\end{align}
where $\sigma^z_i = \pm 1$ indicates whether the current on atom $i$ is flowing CW/CCW with respect to an appropriately oriented local bond. 

The Coulomb interaction between two atoms $i$ and $j$ can be expressed as a quadratic form in the number operators on each atom. 
\begin{align}
  H_{Coulomb} = \sum_{i,j,a,b} V^{ij}_{ab} n_{i,a} n_{j,b}
\end{align}
where $n_{i,a} = c^{\dagger}_{i,a}c_{i,a}$ and $a, b$ run over the four orbitals of each atom. 
Since there are exactly two electrons on each atom, we have the linear constraints $\sum_a n_{i,a} = 2$ so we can express the quadratic form in terms of 3 linearly independent number operators. 

For specificity, consider an atom $i$ lying on a $\hat{x}$ oriented bond. 
Looking down the $\hat{x}$ axis, we index the orbitals
\begin{equation*}
    \tikz[scale=0.5]{
        \draw[->] (0,-2.5) -- (0,2.5) node[right] {$y$};
        \draw[->] (-2.5,0) -- (2.5,0) node[above] {$z$};

        \coordinate (s0) at (1,1);
        \coordinate (s1) at (-1,1);
        \coordinate (s2) at (-1,-1);
        \coordinate (s3) at (1,-1);

        \filldraw (s0) circle (4pt) node[left] {$0$};
        \filldraw (s1) circle (4pt) node[right] {$1$};
        \filldraw (s2) circle (4pt) node[right] {$2$};
        \filldraw (s3) circle (4pt) node[left] {$3$};

        \draw (0,0) circle (1.8);
    }
\end{equation*}
The 3 linearly independent electric moment operators can then be taken as
\begin{align}
    d^z_i    &= d_0 (~~n_{i,0} + n_{i,1} - n_{i,2} - n_{i,3}) \nonumber\\
    d^y_i    &= d_0 (-n_{i,0} + n_{i,1} + n_{i,2} - n_{i,3}) \nonumber\\
    Q^{yz}_i &= Q_0 (~~n_{i,0} - n_{i,1} + n_{i,2} - n_{i,3})
\end{align}
Here, we have introduced dimensionful factors of $d_0$ and $Q_0$ which come from the geometry of the atom and dropped any 1-body terms as these have already been absorbed into the single-atom crystal field Hamiltonian $H_{XF}$.
These operators are adapted to the cubic symmetry of the system and have a particular simple action on the model atomic states: the dipole operators change the momentum of one electron by $\pm \pi/2$ while the quadrupole $Q^{yz}$ shifts it by $\pi$. 

Now let us consider a representative square plaquette lying in the XY plane from $(0,0,0)$ to $(1,1,0)$ with four atoms at $(\pm 1/2, 0)$ and $(0,\pm1/2)$.
Cubic symmetry dictates that among the 9 possible nearest neighbor terms, only 4 are non-zero.
Terms of the type $d^z d^{x}$ and $Q^{xz} d^y$ are disallowed by the mirror reflection $z\to-z$. 
This leaves the orbits of the terms $d^z_1 d^z_2$, $d^y_1 d^x_2$, $Q^{xz}_1 d^z_2$, $Q^{xz}_1 Q^{yz}_2$ under the symmetry of the lattice.

In the usual quantum spin ice hierarchy, $t \gg J \gg V$, we treat $V$ as a perturbation acting on the low energy ice manifold -- the ground states $P_0$ of $H_{XF} + H_{Ice}$. 
In the Schrieffer-Wolf (SW) approach, the perturbative unitary $e^{S}$ with $S = S_1 + S_2 + \cdots$ can be used to rotate operators into the low energy manifold (we follow the notation of \cite{Bravyi2011a}). 
The transverse atomic dipole moment operator of the $XY$ plaquette $p$ is 
\begin{align}
\label{eq:toy_dz_eff_SW}
    d^z_{eff} &= P_0 e^{\hat{S}}(\sum_{i \in p} d^z_i)P_0 
\end{align}
where $\hat{S}$ denotes the superoperator adjoint of $S$, $\hat{S}(X) = [S,X]$. 
While it is straightforward to expand to Eq.~\eqref{eq:toy_dz_eff_SW} to low orders mechanically, the full expressions are not so enlightening.
Rather, let us consider a typical term in the expansion of Eq.~\eqref{eq:toy_dz_eff_SW}. 
This is composed of a string of two-body operators coming from $H_{Coulomb}$, along with appropriate energy denominators of the intermediate states, and a single insertion of $d^z_i$. 
In order to connect between ice states and not vanish when restricted to $P_0$, the momentum on each atom must shift by $\pi$ -- which is accomplished by an even number of dipole operators or an odd number of quadrupole operators. 
This is impossible if the Coulomb potential only includes dipole-dipole or quadrupole-quadrupole terms. 
However, so long as the dipole-quadrupole term, $Q^{xz}d^z$ is available, then terms of the form $P_0 \cdot Q^{zz}_4 Q^{zz}_3 \cdot Q^{zz}_2 d^y_1 \cdot d^z_1 P_0$ (neglecting denominators) do not vanish and we find the \emph{atomic transition dipole mechanism},
\begin{align}
\label{eq:toy_atomic_eff_d}
d^z_{eff} \sim d_0 \frac{V^2}{t J} \mathcal{O}_p 
\end{align}
where the imaginary ring exchange operator
\begin{align}
    \mathcal{O}_p &= i (\sigma^+_0 \sigma^-_1 \sigma^-_2 \sigma^+_3 - \mathrm{h.c.})
\end{align}
is the cubic lattice analogue of Eq.~\eqref{eq:micro_b_polarization}.

\paragraph{Ionic Mechanism---} 
The magnetic atoms in spin ice materials are typically ionized and surrounded by a matrix of compensating, non-magnetic ions. Since the restoring force for ionic displacements are typically smaller than the atomic energy scale $t$, motion of the ions can dominate the electric polarization. 
In the toy model, we can understand this by adding an additional, non-magnetic ion at the center of the square plaquette. 
Symmetry allows the transverse position $z$ of the ion to couple to $d^z$ and $Q^{xz}$ as follows, 
\begin{align}
    H_{ion} &= \frac{p_z^2}{2M} + \frac{1}{2} M \Omega^2 z^2 \\
    &+ V_{d}\, z (d^z_0 + d^z_1 + d^z_2 + d^z_3)  \\
    &+ V_{Q}\, z (-Q^{yz}_0 + Q^{xz}_1 + Q^{yz}_2 - Q^{xz}_3)
\end{align}
where $M$ is the mass of the ion and $\Omega$ is of order the Debye frequency for the optical mode associated with ionic motion. 
Rotating $H_{ion}$ into the ice manifold, it is clear that the $V_{d}$ term rotates into the form in Eq.~\eqref{eq:toy_atomic_eff_d}, and results in a coupling between the ionic polarization and $\vec{O}_p$,
\begin{align}
    d^z_{ion} \sim Z_{ion} \frac{V_{d} d_0}{M \Omega^2} \frac{V^2}{t J} \mathcal{O}_p
\end{align}
This may provide a quantitative enhancement of the effective polarization coupling. 

The quadrupole coupling, $V_{Q}$, also rotates into the ice manifold
\begin{align}
    \sim V_{Q} z Q_0 \frac{V^2}{J^2} \mathcal{O}_p
\end{align}
which directly couples the center ion position to $\mathcal{O}_p$ without needing virtual excitation of the internal atomic level structure at (the typically large) energy scale $t$.
Although this coupling is naively larger, it appears with opposite signs on the 4 neighboring XY plaquettes and thus cancels in the long-wavelength limit. 

\end{document}